# Intrinsic transparent conductors without doping

Xiuwen Zhang[1], Lijun Zhang[1, †], John D. Perkins[2], and Alex Zunger[1, *]

[1]University of Colorado, Boulder, CO 80309, USA
[2]National Renewable Energy Laboratory, Golden, CO 80401, USA

*Corresponding author Email: alex.zunger@colorado.edu

Transparent conductors (TC's) combine the usually contraindicated properties of electrical conductivity with optical transparency and are generally made by starting with a transparent insulator and making it conductive via heavy doping, an approach that generally faces severe 'doping bottlenecks'. We propose a different idea for TC design—starting with a metallic conductor and designing transparency by control of intrinsic interband transitions and intraband plasmonic frequency. We identify the specific design principles for three such prototypical intrinsic TC classes and then search computationally for materials that satisfy them. Remarkably, one of the intrinsic TC, $Ag_3Al_{22}O_{34}$, is predicted also to be a prototype 3D compounds that manifest natural 2D electron gas (2DEG) regions with very high electron density and conductivity.

[†]Present address: College of materials science and Engineering, Jilin University, Changchun 130012, China

The functionality of transparency plus conductivity [1, 2] lies at the center of many technological applications such as solar cell, touch-screen sensors, light emitting diode, electronic papers, infrared or ultraviolet photo detector, smart windows and flat panel display [1-8], yet materials with such seemingly contraindicated properties are difficult to come by. The traditional strategy for searching TC's has followed the path illustrated by the arrow in Fig. 1(a): start from a transparent insulator and find ways to make it conductive by doping it extensively without affecting its optical transparency [1-8]. Successful examples are very few and include Al-doped ZnO and Sn-doped $In_2O_3$ for electron-conducting (*n*-type) TC's [3, 4, 7], as well as hole doped $CuAlO_2$ and La-doped $SrGeO_3$ for hole-conducting (*p*-type) TC's [5, 8]. The limiting factors are rooted in defect physics [9-11] and include difficult to fulfill requirements such as finding wide-gap insulators that can be amply doped without promoting carrier compensation or structural deformations.

In this Letter, we revisit the basic physics design principles needed for transparent conductivity and find that a different, previously overlooked route, illustrated by the arrow in Fig. 1(b), may be possible—start from an opaque conductor that already has plenty of free carriers, then design optical transparency to realize an intrinsic (i.e., without intentional chemical doping) TC. However, not all bulk conductors will do; one needs to search for bulk metals that (a) have a sufficiently broad energy window in their electronic structure either below the Fermi energy $E_F$ (for *n*-type TC) or above $E_F$ (for *p*-type TC), so the interband transitions across the 'energy window' will not obscure optical transparency, and (b) do not have a high plasma frequency ($\omega_p$) [12] so the free carrier reflection will not limit transparency (recall that normal metals such as Al [12] have a high plasma energy of ~15 eV leading to strong reflectivity for most visible light [13]). If one can find metals that satisfy such conditions this would results in the interesting case of metallic conductivity in a transparent and pristine (undoped) crystal. This approach is applicable to *bulk compounds*, and is thus different from the approach of using ultra thin films of metallic materials that are transparent only when they are kept ultra thin [14-17].

The two conditions noted above can appear unusual and indeed materials satisfying them have, to our knowledge, not been deliberately searched before. We focus here on nontraditional metals such as complex oxides and other multinary metal chalcogenides and search computationally for those that satisfy these conditions. We identified the prototype behaviors of intrinsic TC's, discuss their essential properties and find via atomistic, material-dependent electronic structure theory specific compounds that illustrate these prototypes, including $Ag_3Al_{22}O_{34}$, $Ba_3Nb_5O_{15}$ and $Rb_4Nb_{11}O_{30}$. Remarkably, the free electrons predicted to exist in $Ag_3Al_{22}O_{34}$ are found to be spatially organized as *two dimensional electron gas (2DEG), periodically embedded in the 3D compound.*

To study the prototype behaviors of intrinsic transparent conductors, we evaluate their electronic structures, dielectric function and optical properties by the density functional theory (DFT) [18, 19] as well as hybrid functional (HSE06) [20] (see Supplementary sections I and II for details).

***Type-1 intrinsic TC are metals with an isolated intermediate-band:*** The first type (ITC-1) illustrated in Fig. 2 is based on metallic, intermediate-band (IB) materials where the IB is energetically isolated from the bands below and above it and the Fermi energy is located within that band. The particular example shown is based on RbTe in the zincblende crystal structure (Fig. 2(a)) with electronic structure shown in Fig. 2(b).

The *general* required design principles for this prototype (ITC-1) include: (i) the bands above the IB and the bands below IB need to be separated by broad energy windows from the IB (see e.g. Fig. 2(b)) so as to prevent interband transitions by visible light photons. (ii) The energy spacing between different subbands within the IB needs to be smaller than visible light photon energy (Fig. 2(b)) so as to prevent inter-subband transitions with visible-light photon energies. The way the narrowness of the IB helps to defeat inter-subband absorption can also be used in conventional, chemically doped p-type TC's—such narrow bands could lead to high transparency, free of inter-valence-bands transitions. (iii) The carrier density partially filling the IB needs to be in the region that gives a plasma frequency lower than the frequency of visible light but high enough for good conductivity [13] $\sigma = \frac{(\omega_p)^2}{4\pi\gamma}$ where $\gamma$ is the damping coefficient [13] (we use

$\gamma = 0.2$ eV/ℏ analogous to traditional TC's [7]). The need to satisfy such multiple electronic structure functionalities is key to identification of such rare compounds. Fortunately, this is possible, as shown below.

Following these design principles, we focused our attention on the family of I-VI compounds (I = K, Rb and VI = Se, Te) in the zincblende (ZB) structure, the reasons being that such sub-octet I-VI compounds have a partially filled chalcogen *p*-band that is energetically isolated from the alkali ion *s* bands above it and from the chalcogen ion *s* bands below it, thus forming a separate intermediate band (see e.g. Fig. 2(b)). Zincblende RbTe is found to satisfy the conditions for ITC-1 rather well: the bulk optical absorption coefficient (Fig. 2(c)) including plasmonic effect based on Drude model [13] shows nearly zero optical absorption for most visible light. The plasma frequency [13] $\omega_p \sim \sqrt{n_h/m^*}$ (0.56 eV/ℏ) is low enough for transparency due to the large effective mass $m^*$ of IB, but high enough for good conductivity ($0.21 \times 10^3$ S/cm) due to the high hole density ($n_h = 6.71 \times 10^{21}$ cm$^{-3}$). Indeed the simulated optical reflection and transmission spectra for a free standing 1 μm thick slab with optically smooth surfaces (Fig. 2(d)) shows that the sample has very high transmittance (T) and low reflectance (R) for most visible light. Additional I-VI compounds in the ZB structure have analogous ITC-1 properties (such as RbSe, KSe and KTe). However, these compounds are used for illustration purposes only as the ZB structure being an ITC-1 is not in the ground states structure for these compounds (see discussion of thermodynamic aspects in Supplementary section III).

*Type-2 intrinsic TC is an indirect gap semi-metal with large direct band gaps:* The ITC-2 type is based on semimetals having a large *vertical* band gap (assuring optical transparency), yet a zero *indirect band gap*, assuring semi-metallic behavior. The particular example shown is based on compressed silicon in the diamond-like (Fd-3m) structure (Fig. 3(a)) with DFT electronic structure calculations shown in Fig. 3(b).

The *general* required design principles for this prototype (ITC-2) include: (i) large direct gap above the photon energy of most visible light between the $(N_e)^{th}$ and $(N_e+1)^{th}$ bands ($N_e$ is the number of electrons in the primitive cell) so the optical transition between them does not affect the transparency for visible light, and a zero indirect gap. This requires the $(N_e)^{th}$ and $(N_e+1)^{th}$ bands to be highly dispersive and nearly parallel in a portion of the Brillouin zone (e.g. along the Γ-X direction in Fig. 3(b)). (ii) The carrier densities (equal amount of electrons and holes) need to be low enough to achieve a small plasma frequency and a weak optical transition of the electrons (holes) from the $(N_e+1)^{th}$ $((N_e)^{th})$ band to higher (lower) bands.

Diamond-like (Fd-3m) Si at high pressure (50 GPa) is chosen to illustrate the design principles although this structure is not the stable phase for highly compressed silicon (for pressure higher than ~11.2 GPa, Si transforms [21] into the $β$-Sn I4$_1$/amd structure that is an opaque metal with calculated plasma frequency > 9 eV at pressures 0~50 GPa). Within this caveat, compressed Si is found to satisfy the conditions for ITC-2 rather well: the low carrier concentration ($n = 0.08×10^{21}$ cm$^{-3}$) leads to low plasma frequency (0.58 eV/ℏ). Indeed the evaluated absorption coefficient (Fig. 3(c)) is nearly zero for most visible light and very small for infrared light. The transmittance (Fig. 3(d)) is mainly limited by the reflectivity that shows strong oscillatory pattern due to the coherent internal reflections. In actual technological applications, this can be largely mitigated through the use of antireflection coatings or optically rough surfaces.

***Type-3 intrinsic TC is a near-octet metal:*** The third type of intrinsic TC is based on metallic compounds that have a near-octet electronic structure. The particular example shown (Fig. 4(a)) is based on $Ag_3Al_{22}O_{34}$ in the hexagonal P6$_3$/mmc structure [22] (inset of Fig. 4(b)). Considering the standard formal charges of the constituents Al = 3+, Ag = 1+ and O = 2-, the compound $Ag_3Al_{22}O_{34}$ would have 1×3 + 3×22 - 2×34 = +1 nonzero residual valence per formula unit (the *net* physical charge is, however, zero as the nuclear charges compensate the electronic charges). Such near-octet compounds can be made, among other methods, by starting with wide gap octet insulator such as $Ca_{12}Al_{14}O_{33}$ [6] and reducing it to $Ca_{12}Al_{14}O_{32}$ [23], which is a metal. This example, tried previously[25], however, is not really transparent (see Supplementary Fig. S6).

The *general* required design principles for this prototype (ITC-3) include: (i) The $(N_e-δ)^{th}$ and $(N_e-δ+1)^{th}$ bands ($N_e$ is the number of electrons and δ is the residual valence per primitive cell) need to be separated by a large energy window (see e.g. Fig. 4(a)), so the interband optical transition across the energy window does not affect the transparency for visible light. (ii) The carrier density ($n$) and dispersion of the partially filled bands (see

Fig. 4(a)) need to be sufficiently low for low plasma frequency [13] $\omega_p \sim \sqrt{n/m^*}$. (iii) The optical transition of the electrons (holes) from the partially filled bands to the bands above (below) them needs to be weak so as not to adversely affect transparency.

Following the formulated design principles, we inspect a few hundreds ternary oxides in ICSD [22], looking for near-octet residual valence $|\delta| = 1$ (such as $Ag_3Al_{22}O_{34}$ having $\delta = 1$) with low carrier density. We readily identify three candidate ITC-3 materials: $Ag_3Al_{22}O_{34}$ (*n*-type), $Ba_3Nb_5O_{15}$ (*n*-type), and $Rb_4Nb_{11}O_{30}$ (*p*-type). Their thermodynamic stability is demonstrated in Supplementary section V. $Ba_3Nb_5O_{15}$ and $Rb_4Nb_{11}O_{30}$ are found to be stable ground state compounds, whereas $Ag_3Al_{22}O_{34}$ is slightly higher in energy (0.033 eV/atom) than its competing phases ($AgAlO_2$, $Ag$, and $Al_2O_3$). In $Ba_3Nb_5O_{15}$ (see Fig. S10) and $Rb_4Nb_{11}O_{30}$ (see Fig. S11), there are dense energy bands that consist of Nb-*d* and O-*p* states (see local density of states in Fig. S3) near $E_F$. The optical transition between these bands leads to rather strong optical absorption at photon energies 1~2 eV (Figs. S10 and S11). On the contrary, in $Ag_3Al_{22}O_{34}$, there are very few energy bands (Ag-*s*-like, see Fig. S3) near $E_F$, and this sparsity of bands helps to reduce interband optical absorption. Indeed, $Ag_3Al_{22}O_{34}$ is found to satisfy the conditions for ITC-3 rather well: the bulk optical absorption coefficient (Fig. 4(b)) shows nearly zero absorption for most visible light, except the absorption peak near 3 eV for *z*-polarized light (see green curve in Fig. 4(b)) originating from the interband optical transitions from the partially filled band below $E_F$ to the bands above $E_F$. The interplay between medium electron density ($1.58 \times 10^{21}$ cm$^{-3}$) and medium band dispersion (see Fig. 4(a)) leads to small plasma frequencies ($\omega_p^{xx} = \omega_p^{yy} = 1.14$ eV). The calculated transmission spectrum of 1 μm thick slab (Figs. 4(c) and 4(d)) shows an overall transparency of ~70%.

**Natural 2-dimensional electron gas (2DEG) forming in a 3D bulk compound:** Interestingly, we find that $Ag_3Al_{22}O_{34}$ in the hexagonal P6$_3$/mmc crystal structure has very high in-plane (*xy*-plane) conductivity ($\sigma^{xx} = \frac{(\omega_p^{xx})^2}{4\pi\gamma} = 0.88 \times 10^3$ S/cm) but zero out-of-plane conductivity, *forming a 2DEG in a bulk compound* without the need for Molecular Beam Epitaxy (MBE) synthesized heterostructures with designed modulation doping [24, 25]. Another possible system [26] $Ca_2N$, does not have a truly 2D electron layer as according to DFT it has rather high out-of-plane conductivity (see Supplementary section VII) in comparison to the pure 2D conductivity in $Ag_3Al_{22}O_{34}$ (Fig. 4). To demonstrate the distribution of carriers in intrinsic TC $Ag_3Al_{22}O_{34}$, we summed the charge density set up by wavefunctions in the energy region indicated by the yellow shading in Fig. 4(a) (between $E_F$-1 to $E_F$ eV) and obtain the real space electron density shown in Fig. 5(a). We see that the two dimensional electron gas is confined primarily to the Ag-O layers and separated by the Al-O barriers (Fig. 5(a)). The carrier density (Fig. 5(b)) of the 2DEG in the lower Ag-O region is about two times larger than that in the upper Ag-O layer, proportional to the number of Ag atoms (see inset of Fig. 4(b)). *It is remarkable that the carrier density of the 2DEG in the lower Ag-O layer is as high as $10^{16}$ cm$^{-2}$ (Fig. 5(b))—much higher density than the carrier density achieved in 2DEGs produced in MBE heterostructures ($10^{11}$~$10^{12}$ cm$^{-2}$ in semiconductor heterostructures and $10^{13}$~$10^{14}$ cm$^{-2}$ in oxide interfaces [25]).* We note that the spacing (~1 nm) and periodicity (~2 nm) of the alternating higher versus lower density 2DEGs are rather small, thus the 2DEGs can couple with each other. This type of periodic high carrier-density 2DEGs in 3D compounds offers the route to study the mesoscopic collective effects of interacting periodic 2DEGs. The 2D conductivity of 2DEGs is also preferred for the high-performance 2D TC layers in devices for avoiding carrier scattering at the surfaces of TC layers.

***Conclusions***: The strategy of designing TC's without deliberate doping (Fig. 1(b)) is a particular case of a broader approach of inverse design [10, 27]—starting from physics based 'design principles' (DP), then constructing the 'design metrics' (DM) that are computable quantities that embody the physics of the DP, followed by extensive search of materials that score highly on the DM scale, leading to the identification of specific, few 'best of class' materials. Here, we extend the initial step of inverse design by revisiting the basic design principles of a selected functionality, leading us to the potentially overlooked prototypes of functional materials, such as the bulk compounds that support free carriers without extrinsic doping while maintaining transparency predicted in this study. Avoidance of deliberate doping (compare Fig. 1(a) with 1(b)) may circumvent structural defects and could thus simplify the manufacturing techniques compared to processes that rely on heavy, and often non-equilibrium doping. Indeed, more extended search of these functionalities, in parallel with stability and growability calculations (exemplified by Supplementary Figs. S7-S9) along with experimental scrutiny of such results might well be the way to accelerated discovery of functional materials.

**Acknowledgements**

This work was supported by the U.S. Department of Energy, Office of Science, Basic Energy Sciences. We thank Liping Yu and Giancarlo Trimarchi for helpful discussions.

**References**

[1]  D.S. Ginley, H. Hosono, D.C. Paine, Handbook of Transparent Conductors, (Springer Science & Business Media, 2010).

[2]  A.V. Moholkar, Transparent Conductors, (LAP Lambert Academic Publishing, 2011).

[3] K. Wasa, S. Hayakawa, T. Hada, Electrical and Optical Properties of Sputtered n-p ZnO–Si Heterojunctions, Jpn. J. Appl. Phys. 10, 1732 (1971).

[4] I. Hamberg, A. Hjortsberg, C.G. Granqvist, High quality transparent heat reflectors of reactively evaporated indium tin oxide, Appl. Phys. Lett. 40, 362 (1982).

[5] H. Kawazoe, M. Yasukawa, H. Hyodo, M. Kurita, H. Yanagi, H. Hosono, P-type electrical conduction in transparent thin films of $CuAlO_2$, Nature 389, 939 (1997).

[6] K. Hayashi, S. Matsuishi, T. Kamiya, M. Hirano, H. Hosono, Light-induced conversion of an insulating refractory oxide into a persistent electronic conductor, Nature 419, 462 (2002).

[7] G.V. Naik, J. Kim, A. Boltasseva, Oxides and nitrides as alternative plasmonic materials in the optical range, Optical Materials Express 1, 1090 (2011).

[8] H. Mizoguchi, T. Kamiya, S. Matsuishi, H. Hosono, A germanate transparent conductive oxide, Nature Communications 2, 470 (2011).

[9] Ç. Kılıç, A. Zunger, Origins of Coexistence of Conductivity and Transparency in $SnO_2$, Phys. Rev. Lett. 88, 095501 (2002).

[10] T.R. Paudel, A. Zakutayev, S. Lany, M. d'Avezac, A. Zunger, Doping Rules and Doping Prototypes in $A_2BO_4$ Spinel Oxides, Adv. Funct. Mater. 21, 4493 (2011).

[11] G. Hautier, A. Miglio, G. Ceder, G.-M. Rignanese, X. Gonze, Identification and design principles of low hole effective mass p-type transparent conducting oxides, Nat. Commun. **4**, 2292 (2013).

[12] E.D. Palik, Handbook of Optical Constants of Solids, (Academic Press, Boston, 1998).

[13] P. Drude, Zur Elektronentheorie der Metalle, Annalen der Physik **306**, 566 (1900).

[14] K.S. Kim, Y. Zhao, H. Jang, S.Y. Lee, J.M. Kim, K.S. Kim, J.-H. Ahn, P. Kim, J.-Y. Choi, B.H. Hong, Large-scale pattern growth of graphene films for stretchable transparent electrodes, Nature **457**, 706 (2009).

[15] T. Ohsawa, J. Okubo, T. Suzuki, H. Kumigashira, M. Oshima, T. Hitosugi, An n-Type Transparent Conducting Oxide: $Nb_{12}O_{29}$, J. Phys. Chem. C **115**, 16625 (2011).

[16] J. van de Groep, P. Spinelli, A. Polman, Transparent Conducting Silver Nanowire Networks, Nano Lett. **12**, 3138 (2012).

[17] X. Meng, D. Liu, X. Dai, H. Pan, X. Wen, L. Zuo, G. Qin, Novel stable hard transparent conductors in $TiO_2$-TiC system: Design materials from scratch, Sci. Rep. **4**, 7503 (2014).

[18] J.P. Perdew, K. Burke, M. Ernzerhof, Generalized Gradient Approximation Made Simple, Phys. Rev. Lett. **77**, 3865 (1996).

[19] G. Kresse, D. Joubert, From ultrasoft pseudopotentials to the projector augmented-wave method, Phys. Rev. B **59**, 1758 (1999).

[20] J. Heyd, G.E. Scuseria, M. Ernzerhof, Erratum: "Hybrid functionals based on a screened Coulomb potential", J. Chem. Phys. **124**, 219906 (2006).

[21] J.Z. Hu, I.L. Spain, Phases of silicon at high pressure, Solid State Comm. **51**, 263 (1984).

[22] ICSD, Inorganic Crystal Structure Database; Fachinformationszentrum Karlsruhe: Karlsruhe, (Germany, 2006).

[23] S. Matsuishi, Y. Toda, M. Miyakawa, K. Hayashi, T. Kamiya, M. Hirano, I. Tanaka, H. Hosono, High-Density Electron Anions in a Nanoporous Single Crystal: [Ca24Al28O64]4+(4e-), Science **301**, 626 (2003).

[24] R. Dingle, H.L. Störmer, A.C. Gossard, W. Wiegmann, Electron mobilities in modulation-doped semiconductor heterojunction superlattices, Appl. Phys. Lett. **33**, 665 (1978).

[25] J. Mannhart, D.G. Schlom, Oxide Interfaces—An Opportunity for Electronics, Science **327**, 1607 (2010).

[26] K. Lee, S.W. Kim, Y. Toda, S. Matsuishi, H. Hosono, Dicalcium nitride as a two-dimensional electride with an anionic electron layer, Nature **494**, 336 (2013).

[27] A. Franceschetti, A. Zunger, The inverse band-structure problem of finding an atomic configuration with given electronic properties, Nat. Commun. **402**, 60 (1999).

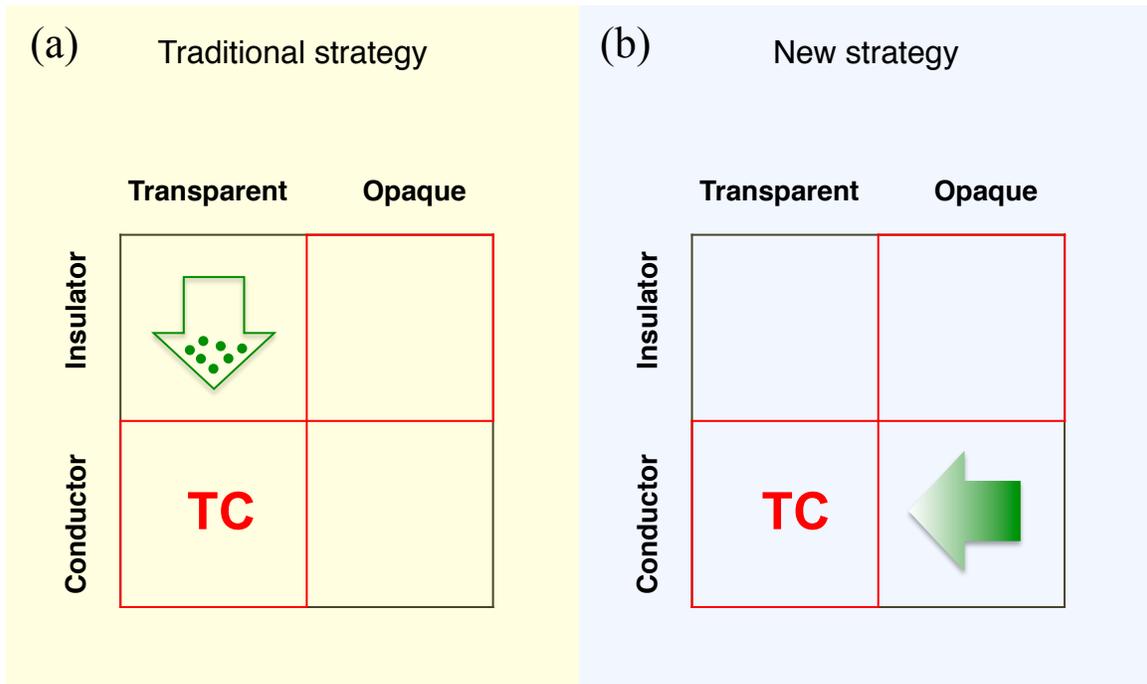

FIG. 1. (a) The traditional strategy for designing bulk transparent conductors that starts from a wide-gap insulator and finds ways to make it conductive by extensive doping without affecting its crystal structure or optical transparency. (b) The new strategy that starts from a metal that already has plenty of free carriers and designs optical transparency to realize an intrinsic (i.e., without intentional chemical doping) TC.

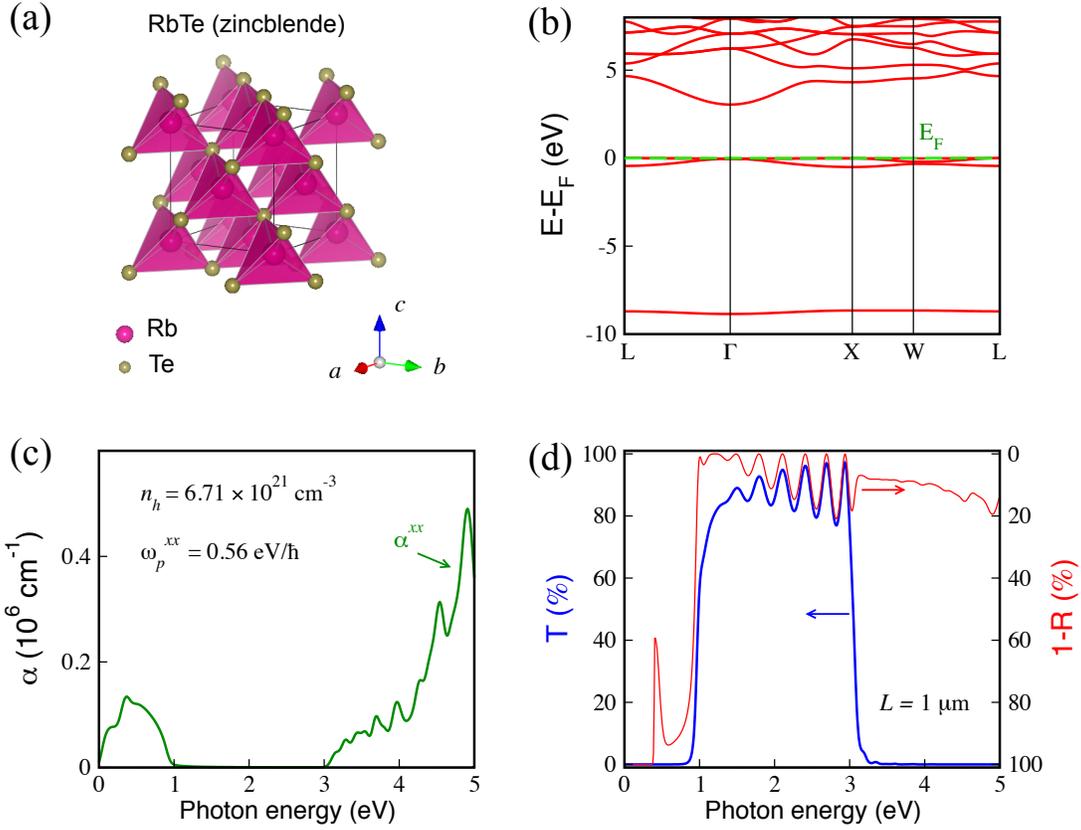

FIG. 2. (a) Crystal structure, (b) band structure and (c) absorption coefficient ($\alpha^{xx} = \alpha^{yy} = \alpha^{zz}$) of zincblende (F-43m) RbTe as an example of type-1 intrinsic TC from DFT. The carrier (hole) concentration ($n_h$) and plasma frequency ($\omega_p^{xx} = \omega_p^{yy} = \omega_p^{zz}$) are given in (c). The z-axis is chosen along the [001] direction of the cubic lattice. (d) Transmission and reflection spectra of a free standing 1 μm thick RbTe (F-43m) slab with optically smooth surfaces.

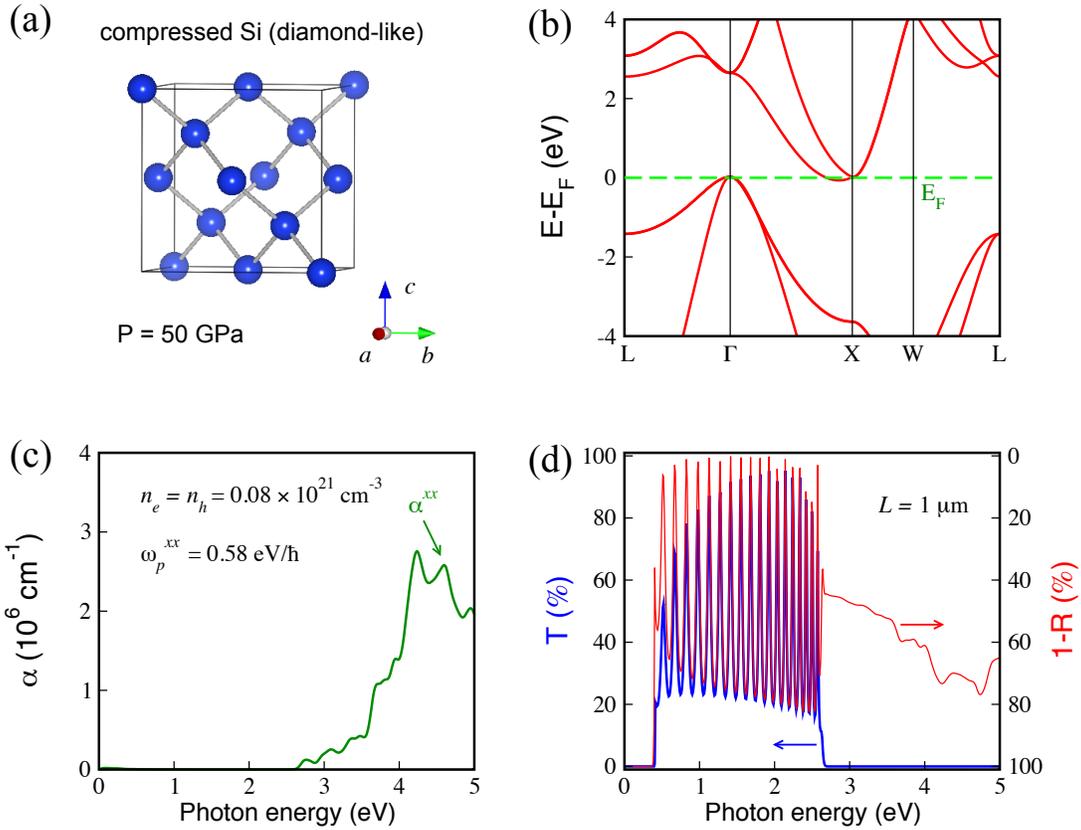

FIG. 3. (a) Crystal structure, (b) band structure and (c) absorption coefficient ($\alpha^{xx} = \alpha^{yy} = \alpha^{zz}$) of compressed diamond-like (Fd-3m) Si under high pressure (P = 50 GPa) as an example of type-2 intrinsic TC from DFT. The carrier (electron and hole) concentration ($n_e = n_h$) and plasma frequency ($\omega_p^{xx} = \omega_p^{yy} = \omega_p^{zz}$) are given in c. The z-axis is chosen along the [001] direction of the cubic lattice. (d) Transmission and reflection spectra of a free standing 1 μm thick compressed Si (Fd-3m) slab with optically smooth surfaces.

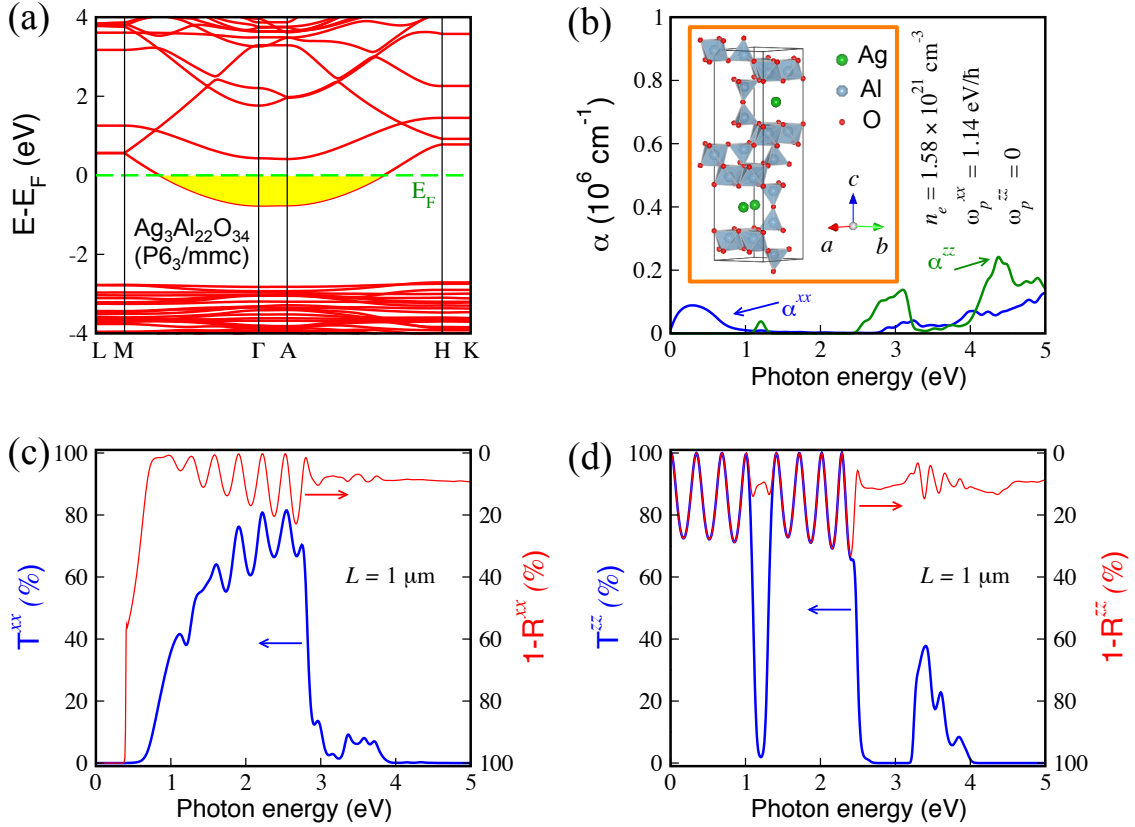

FIG. 4. (a) Band structure, (b) absorption coefficient ($\alpha^{xx} = \alpha^{yy}; \alpha^{zz}$) (inset: crystal structure) of $Ag_3Al_{22}O_{34}$ ($P6_3/mmc$) as an example of type-3 intrinsic TC from DFT. The yellow shading in (a) illustrates the electrons filling the band just below $E_F$. The carrier (electron) concentration ($n_e$) and plasma frequency ($\omega_p^{xx} = \omega_p^{yy}; \omega_p^{zz}$) are given in (b). The $z$-axis is chosen along the [0001] direction of the hexagonal lattice. (c)-(d) Transmission and reflection spectra of a free standing 1 μm thick $Ag_3Al_{22}O_{34}$ slab perpendicular to $x$ direction (c) or $z$ direction (d) with optically smooth surfaces.

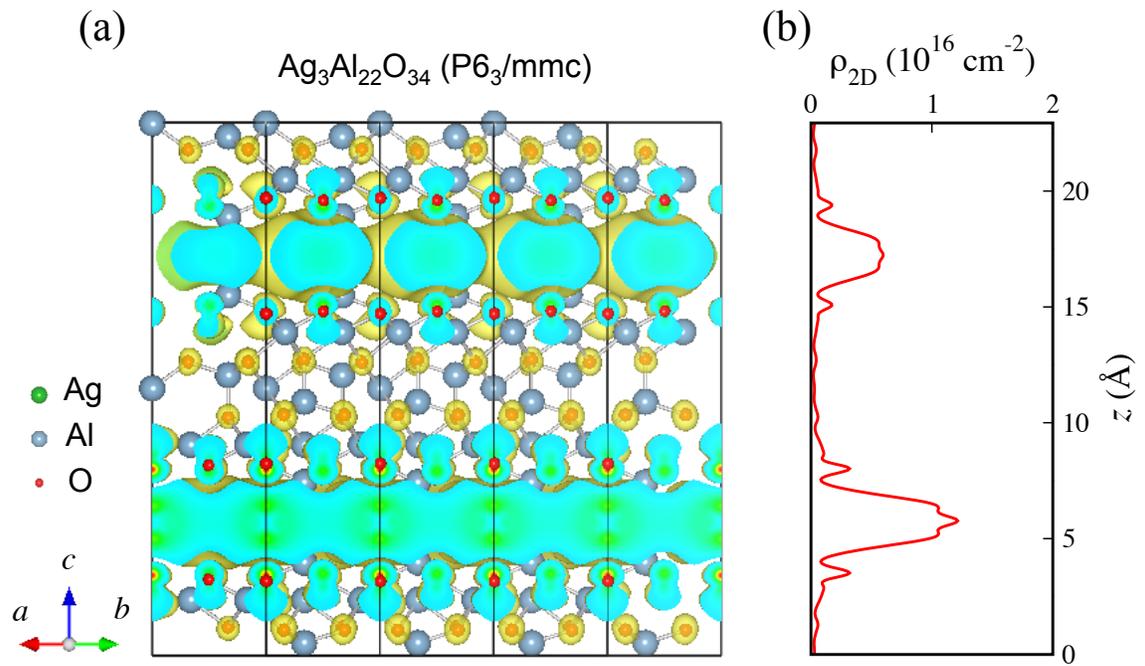

FIG. 5. (a) Real space electron density (isosurface $0.5 \times 10^{21}$ cm$^{-3}$) of Ag$_3$Al$_{22}$O$_{34}$ (P6$_3$/mmc). (b) Two-dimensional carrier density in $xy$-plane as a function of the position $z$.